\documentclass{aa}  

\usepackage{graphicx}
\usepackage{txfonts}
\usepackage{xcolor}
\usepackage[
    locale=UK,
    per-mode=fraction,
    output-decimal-marker={{.}},
    separate-uncertainty=true,
    range-phrase={{ bis }},
]{siunitx} 
\usepackage[breaklinks=true]{hyperref}

\definecolor{ochre}{rgb}{0.8, 0.47, 0.13}

\definecolor{lightgreen}{HTML}{CCFF99}

\begin{document}

   \title{GRB 210619B optical afterglow polarization}

   \author{N. Mandarakas
          \inst{1,2}
           \and D. Blinov\inst{1,2}  \and D. R. Aguilera-Dena\inst{1} \and S. Romanopoulos\inst{1,2} \and V. Pavlidou\inst{1,2} \and K.~Tassis\inst{1,2} \and J.~Antoniadis\inst{1,3} \and S. Kiehlmann\inst{1,2} \and A.~Lychoudis\inst{2} \and L. F. Tsemperof Kataivatis\inst{2} }

   \institute{Institute of Astrophysics, Foundation for Research and Technology - Hellas, Voutes, 70013 Heraklion, Greece\\
              \email{nmandarakas@physics.uoc.gr}
         \and
             Department of Physics, University of Crete, 70013, Heraklion, Greece
        \and            
        Max-Planck-Institut f\"{u}r Radioastronomie, Auf dem H\"{u}gel 69, DE-53121 Bonn, Germany
           }

   \date{Received XXX; accepted March 25, 1821}

 
  \abstract
   {}
   {We report on the follow-up of the extremely bright long gamma-ray burst GRB~210619B with optical polarimetry.}
   {We conducted optopolarimetric observations of the optical afterglow of GRB~210619B in the SDSS-r band in the time window $\sim 5967 - 8245$ seconds after the burst, using the RoboPol instrument at the Skinakas observatory.}
   {We report a $5\,\sigma$ detection of polarization $P=1.5\pm0.3$ at polarization angle $EVPA=8\pm6^\circ$. We find that during our observations the polarization is likely constant. These values are corrected for polarization induced by the interstellar medium of the Milky Way and host-induced polarization is likely negligible. Thus the polarization we quote is intrinsic to the GRB afterglow.}
   {}


   \maketitle
%

\section{Introduction}
Gamma-ray bursts (GRBs) are the most energetic electromagnetic astrophysical phenomena known today. They are typically discovered and classified by their characteristic fast-rising, luminous emission of gamma-ray photons, the so-called prompt emission phase, and are typically followed by a multiband emission, known as afterglow.
Depending on the time during which most gamma-ray photons are detected, GRBs are divided into two categories: short GRBs, which typically last < 2 seconds, and long GRBs, which last $\sim30$ seconds \citep{Kouveliotou1993}. Recently, short GRBs have been observed to form in association with kilonovae and gravitational wave emission, suggesting that they originate in compact object mergers \citep[e.g.][]{Narayan1992,Abbott2017B,Abbott2017A,Makhathini2021}. Long GRBs are theorized to occur after the collapse of a massive star into a black hole \citep[e.g.][]{Woosley1993}, or a magnetar \citep[e.g.][]{Usov1994}, and are often associated with supernovae \citep[e.g][]{Hjorth2003}. GRB afterglows are usually attributed to the interaction between an ultra-relativistic jet that is launched during the formation of the compact object, and the circumburst medium. The observed multiwavelength synchrotron emission is thought to be due to the propagation of two shocks, a forward shock and a reverse shock, with the latter dominating at early times \citep{Piran1999,Piran2004}. If an ordered magnetic field is present in the ejecta, the reverse shock can be highly polarized \citep[e.g.][]{Granot2003}. The polarization of the forward shock depends on the morphology and intensity of the circumburst magnetic field \citep{Uehara2012}.

The exact emission mechanisms, geometry, and physical properties of the emission region in GRBs are currently not well understood. Polarimetric observations of GRBs, both in the prompt phase and during the afterglow, can potentially reveal some of their unknown properties. 
Polarization in the prompt phase of the GRBs in the $\gamma$-ray band is expected to arise due to synchrotron radiation \citep[e.g.][]{Granot2003,Waxman2003,Metzger2011}, inverse Compton \citep[e.g.][]{Shaviv1995}, or fragmented fireballs \citep[e.g.][]{Lazzati2009}, and the expected levels of polarization are in the order of several tens of percent, depending on the model, while observations are in agreement with the predictions (see \cite{Covino2016} for a review). However, recent results by \cite{Kole2020}, reveal cases where the polarization of the prompt phase in the 50-500 keV energy range can be as low as zero.

There have been several polarimetric measurements of optical GRB afterglows. Most late GRB optical afterglows display polarization degree values of a few percent \citep{Covino2016}. However, early-time optical afterglows systematically display higher values of polarization (tens of percent), which has been attributed to ordered magnetic fields within the jets, strong enough not to be distorted by the reverse shock \citep{Deng2017}. Whether magnetic fields around GRBs are mostly ordered or random can have an observable effect on their polarisation, related mostly to the variations of polarization angle in time \citep{Teboul2021}.

There are some notable examples of single-epoch optopolarimetric measurements of GRB afterglows. For example, \cite{Steele2009} measured the polarization of the early afterglow of GRB~090102 at $10.1\pm1.3\%$. \cite{Uehara2012} presented a polarization measurement of GRB~091208B, 149–706 seconds after trigger at the level of $10.4\pm2.5\%$. 

Time-resolved measurements could provide better insights into the processes following a GRB. For example, \cite{Mundell2013} observed the optopolarimetric evolution of GRB~120308A and reported a polarization degree of $28\pm4$, decreasing to $16\substack{+5 \\ -4}\%$, with the polarization angle remaining constant during the observations. 
Recently, \cite{Shrestha2022a} followed GRB~191016A 3987–7687 seconds after the burst and reported evidence of polarization in all phases, with a peak of $14.6\pm7.2\%$ , which coincides with the start of the flattening of the light curve. 
\cite{Shrestha2022} present time-resolved optopolarimetric measurements of several GRB afterglows, finding evidence of polarization in two of them, one of them being GRB~191016A as discussed above.

\subsection{GRB~210619B}
In this paper we report on the optical afterglow polarization of long gamma-ray burst GRB~210619B, $\sim 5967-8245$ seconds post-burst. On June 19, 2021, at 23:59:25 UT, the \textit{Swift} Burst Alert Telescope (BAT) triggered and located GRB~210619B \citep{Davanzo2021}. The GRB was also detected by the \textit{GECAM} \citep{Zhao2021}, the \textit{Konus-Wind} experiment \citep{Svinkin2021} and the \textit{Fermi} Gamma-Ray Burst Monitor \citep{Poolakkil2021}. \cite{Postigo2021} observed the optical afterglow with OSIRIS on the 10.4 m GTC telescope at Roque de los Muchachos Observatory and measured its redshift $z=1.937$. They also detected an intervening system at $z=1.095$. \cite{Atteia2021} noted the probability of lensed visible afterglow in the following months due to this intervening system. 

\cite{Oganesyan2021} combined multi-filter optical observations together with X-ray and $\gamma$-ray data to model the emission of the GRB. They disentangled the contributions of the reverse and forward shocks and argued that the GRB multiwavelength emission is produced by a narrow highly magnetized jet propagating in a sparse environment, with an approximate jet-break time at $\sim 10^4$ seconds. Comparison between the optical and X-ray data shows evidence of a secondary component of radiation in the jet wings.

\section{Observations and data reduction}

\subsection{GRB observations}\label{grb_obs}

We conducted optical polarimetric observations of the afterglow of GRB~210619B with the RoboPol instrument which is mounted on the 1.3 m telescope at the Skinakas Observatory in Crete\footnote{https://skinakas.physics.uoc.gr/}. RoboPol is a four-channel polarimeter without rotating parts that can measure the linear Stokes parameters $I$, $q=Q/I$, and $u=U/I$ in a single exposure \citep{Ramaprakash2019}. The instrument splits the incoming light in four beams, corresponding to four orthogonal polarization directions, and projects four spots on the CCD in a cross-shape for each of the imaged sources.
It is optimized for a single source placed in the center of the field of view by using a mask in the telescope focal plane, which is designed to significantly lower the background and prevent overlapping between sources.
Observations with RoboPol are performed using an automated pipeline \citep{King2014b}. In case of targets-of-opportunity, like GRBs, the automated system informs the observers and provides the coordinates of the event.
For example, RoboPol was used to observe the optical afterglow of GRB~131030A, where \cite{King2014a} measured a constant linear polarization value of $p=2.1\pm1.6\%$ throughout the observations.

Following the trigger of GRB~210619B, regular observations were interrupted and the telescope was pointed towards the GRB location. We began taking exposures in the SDSS-r band at  1:37:12.00 UT, June 20, 2021, $\sim5867$ seconds after the trigger ($\sim1998$ seconds after the burst in the source frame). At the time of observations, it was already morning astronomical twillight. Therefore, we observed the GRB afterglow by taking a series of 200-second exposures until the background was too high to allow for more observations. This resulted in 11 200-second exposures. For the first five of them we were able to confirm that they are likely not affected by the polarization of the morning sky (see Sec.~\ref{sec:results}). Data reduction and calibration were performed using the standard RoboPol pipeline \citep{King2014b,Panopoulou2015,Blinov2021} and the RoboPol instrument model.

\subsection{Interstellar polarization}
\begin{figure}
 \centering
 \includegraphics[width=0.48\textwidth]{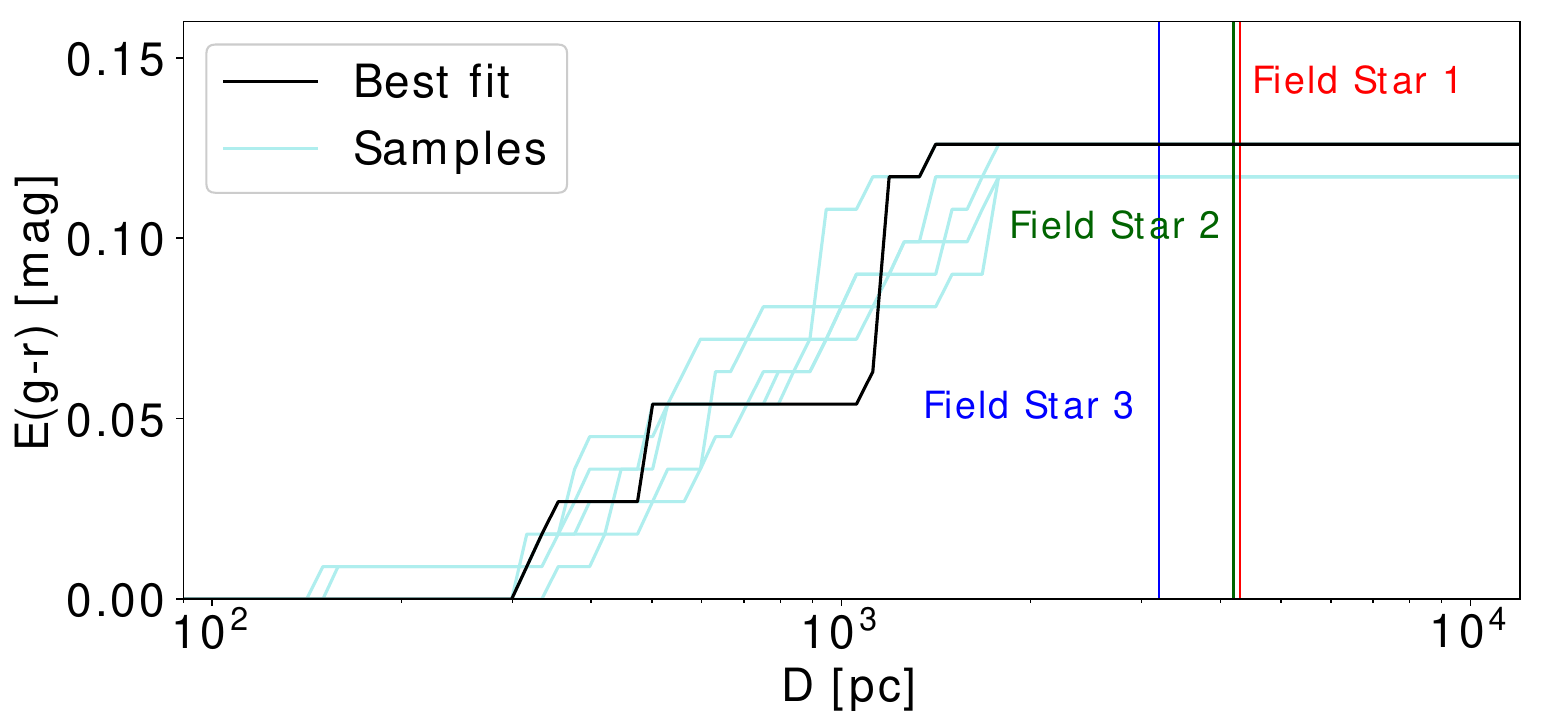}\caption{Reddening ($g-r$) as a function of distance along the line of sight of the GRB, as compiled by \cite{Green2019}. The vertical lines denote the distance of the three field stars used for ISP estimation. The black line represents the maximum-probability density estimate of the distance-reddening curve.}
 \label{fig:Egr}
\end{figure}

\begin{figure}
 \centering
 \includegraphics[width=0.49\textwidth]{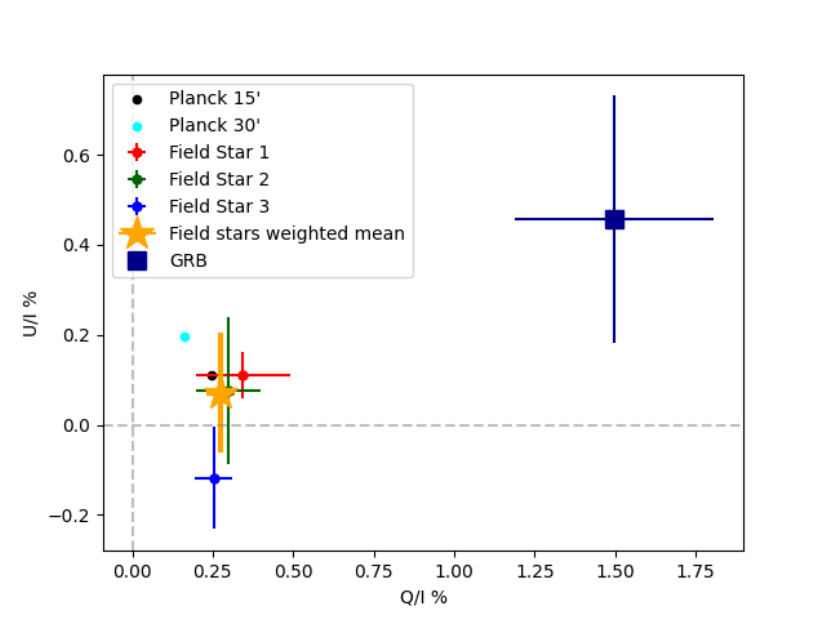}
\caption{Relative Stokes parameters of the measured field stars together with the values obtained by Planck for the line-of-sight of the GRB. We also show the weighted mean values of the first five GRB measurements before correcting for the ISP to higlight the significance of the measurement.}
 \label{fig:ISP}
\end{figure}

The observed polarization of any object is a result of the Milky Way interstellar polarization (ISP) added to the intrinsic one. To account for and correct for the ISP, we observed three field stars in the mask of RoboPol. ISP is produced by the same dust that is responsible for extinction; therefore, stars that are more affected by extinction are expected to provide a better estimate of the polarization fraction induced by the interstellar medium (ISM). We used the 3D dust map compiled by \cite{Green2019} to probe the galactic extinction at different distances from the line of sight of the GRB. We chose bright stars in the field of the GRB that are expected to have maximum extinction (Fig.~\ref{fig:Egr}), and therefore their polarization should reflect, as accurately as possible, the polarization induced by the ISM, provided they are intrinsically zero-polarized. This is a fair assumption as most stars do not have intrinsic polarization, with few exceptions, such as magnetic stars, evolved stars or stars surrounded by a dusty disk \citep{Fadeyev2007,Clarke2010}. In fact, if a peculiar star was present, it would show up as an outlier from the rest in the $q-u$ plot (Fig.~\ref{fig:ISP}). The extinction $E(B-V)$ towards the line of sight of the GRB is $0.12-0.14\,mag$ (depending on the choice of equation for unit conversion between \cite{Green2019} units and $E(B-V)$ - see \cite{dustmaps}). The upper limit for the polarization induced by dust alignment in the ISM is $13\%\times E(B-V)$ \citep{Panopoulou2019,Planck2020}. Based on this estimate, the expected maximum ISM-induced polarization in the line of sight is $1.6-1.8\%$. However, field stars in the same line of sight have an average polarization value of $P=0.26\pm0.05\%$. This could be due to the presence of multiple dust clouds in the line of sight which are permeated by magnetic fields with different orientations. This configuration would align dust grains in different directions in each cloud and thus give rise to depolarization of the light of distant stars \citep[e.g.][]{Tassis2018}. The steps seen in the reddening plot in Fig.~\ref{fig:Egr} hint in such a scenario.

Another way to estimate the ISP, is by using the polarized thermal emission map provided by \cite{Planck2015XIX}. Dust grains absorb optical light, which is preferentially polarized parallel to their major axis. Therefore, in the case where dust is ordered (e.g. due to the presence of a magnetic field), light would appear preferentially polarized in the direction perpendicular to the dust grains. The light absorbed by dust is re-emitted in the far-infrared, with a polarization along the major axes of the grains. It follows from this that thermal emission is expected to be polarized in the plane perpendicular to the optical polarization produced by the ISM. \cite{Planck2015XXI} studied the polarization of 206 stars in the sub-millimeter and optical ranges and provided the correlation of the Stokes parameters $q=Q/I$, and $u=U/I$ between the two bands. The correlation is clear and robust. Thus, the optical ISP can be directly inferred by the submillimeter measurements of \cite{Planck2015XIX} on the same line of sight. We derive the optical ISP using Planck data for the region centered in the location of the GRB with two different resolutions, 30$\arcmin$ and 15$\arcmin$.

We show the measurements of the field stars together with the optical ISP derived by the Planck measurements in Fig.~\ref{fig:ISP}. For comparison, we also plot the weighted mean of the first five GRB measurements on the same plot - not corrected for the ISM contribution (see Sect~\ref{sec:results} for more details). Information on the ISP and mean GRB measurements is presented in Table~\ref{table:fs}.
The weighted mean of the field stars and the ISP value obtained by the Planck 15' measurement are well consistent with each other. Since we only have three field stars for the ISP estimation, we decided to use the Planck value, since the Planck beam  averages over a dust column with a larger cross section, which gives higher SNR. The field stars served as an additional confirmation of the validity of Planck data in the line of sight of interest. We favor the 15$\arcmin$ over the 30$\arcmin$ resolution, since it gives information about the ISP on more localized scales and should better reflect the true value of the ISP in the line of sight of the GRB. Finally, \cite{Skalidis2019} have shown that using Planck submillimeter polarization data in such a scale to infer the optical ISP is a reasonable choice that provides accurate and high-SNR measurements.

In the case when the object is extragalactic, such as GRBs, host galaxy ISM could also potentially polarize the observed light. Host-induced polarization is proportional to the amount of dust present around the location of the burst and can be probed by the reddening of the GRB. For our case, \cite{Oganesyan2021} find that the host galaxy reddening is negligible, and therefore the host-induced polarization should be negligible as well.

The final measurement and uncertainties of Stokes $q=Q/I$ for each of the exposures is simply:

\begin{equation}
    q = q^{measured} - q^{instrumental} -q^{ISP}
\end{equation}

\begin{equation}
    \sigma_q = \sqrt{\left(\sigma_q^{measured}\right)^2 + \left(\sigma_q^{instrumental}\right)^2 + \left(\sigma_q^{ISP}\right)^2}
\end{equation}
and similarly for Stokes $u=U/I$.

\begin{table*}
\caption{Field stars polarization corrected for instrumental polarization and their weighted mean together with the optical ISP derived from Planck using two different resolutions, 30$\arcmin$ and 15$\arcmin$, as well as the weighted mean and errors on the mean of the first five GRB measurements.}            
\label{table:fs}      
\centering          
\begin{tabular}{c c c c l S[table-format=-2.2,table-figures-uncertainty=2] S[table-format=-2.2,table-figures-uncertainty=2]}     
\hline\hline       
star ID & RA  & Dec & Dist. & r       & $Q/I$ & $U/I$ \\
        & deg & deg & kpc   & mag     &   \%  & \%  \\
\hline                    
  Field Star 1 & 319.7258658 & 33.8492013 & 4.3 & 13.85 & 0.34\pm0.15 & 0.11\pm0.05 \\
  Field Star 2 & 319.7152985 & 33.8603076 & 4.2 & 13.90 & 0.30\pm0.10 & 0.08\pm0.16 \\
  Field Star 3 & 319.7351566 & 33.8638350 & 3.2 & 15.7  & 0.25\pm0.06 & -0.12\pm0.11 \\
 \hline 
  Weighted Mean & -- & -- & -- & --  & 0.25\pm0.04 & 0.07\pm0.13 \\
\hline                  
Planck 30$\arcmin$ & 319.71831 & 33.85044 & -- & --  & 0.16\pm0.02 & 0.19\pm0.02 \\
Planck 15$\arcmin$ & 319.71831 & 33.85044 & -- & --  & 0.25\pm0.02 & 0.11\pm0.02 \\
\hline 
GRB & 319.71831 & 33.85044 & -- & --  & 1.50\pm0.31 & 0.46\pm0.28 \\
\hline
\end{tabular}
\end{table*}

\section{Results \& Discussion}\label{sec:results}
                
\begin{figure}
 \centering
 \includegraphics[width=0.48\textwidth]{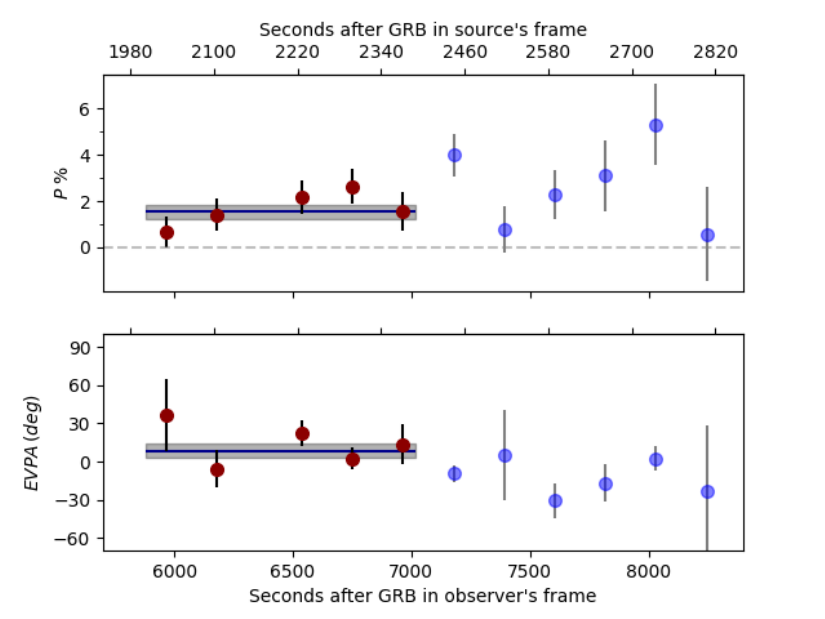}
\caption{Debiased fractional polarization (top) and optical polarization plane direction (bottom) as a function of time. Time is given in both the observer's and the source's frame.Red points are the measurements likely not affected by the morning twilight (see Sec~\ref{sec:results}). The blue points are those most likely affected by twilight. Errorbars correspond to $1\,\sigma$ confidence level. The horizontal blue line and the width of the shaded region correspond to the weighted mean and the $1\,\sigma$ error of the mean of the first five measurements.}
 \label{fig:results}
\end{figure}

\begin{figure*}
   \centering
   \begin{tabular}{cc}
   \includegraphics[trim={0.0cm 0.0cm 0.45cm 0.0cm},clip,width=0.45\textwidth]{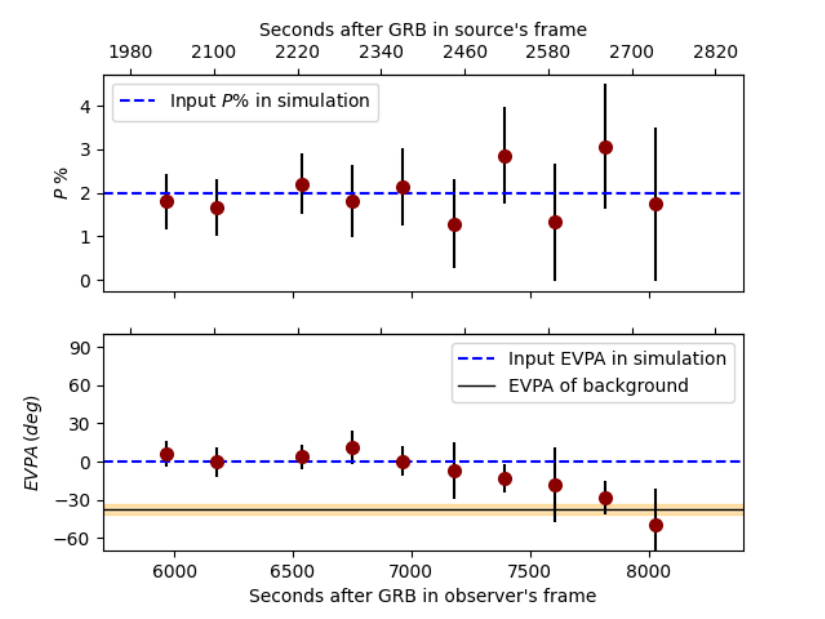}& \vline
   \includegraphics[trim={0.0cm 0.0cm 0.45cm 0.0cm},clip,width=0.45\textwidth]{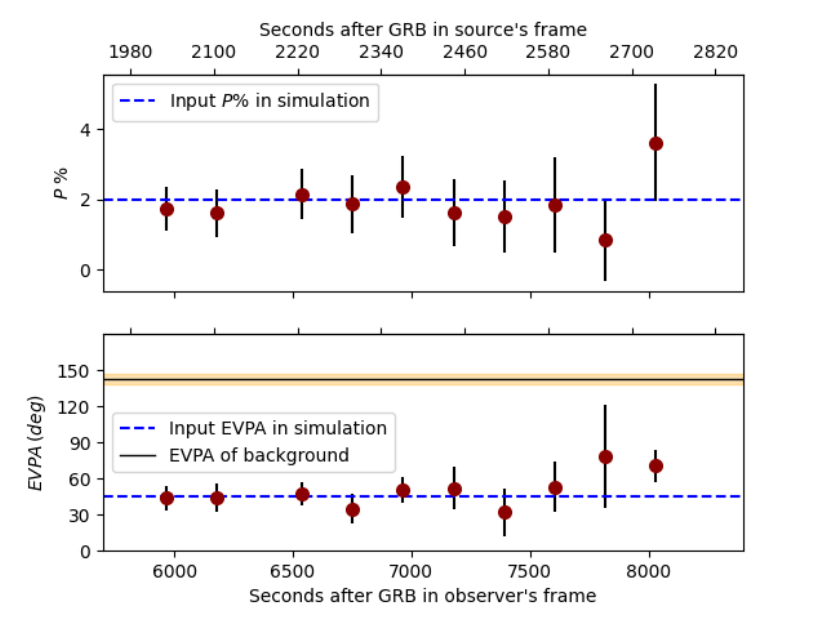}\\
   \hline
   \includegraphics[trim={0.0cm 0.0cm 0.45cm 0.0cm},clip,width=0.45\textwidth]{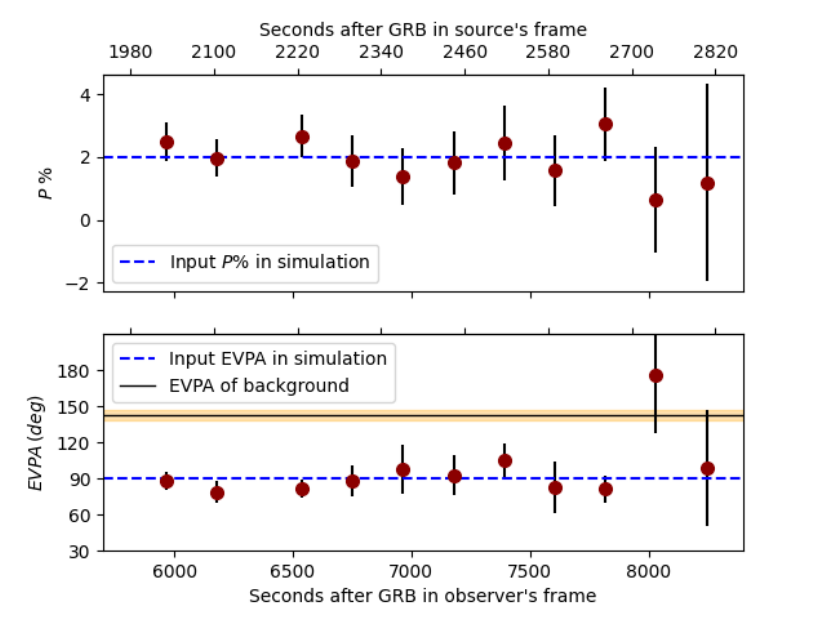}& \vline
   \includegraphics[trim={0.0cm 0.0cm 0.45cm 0.0cm},clip,width=0.45\textwidth]{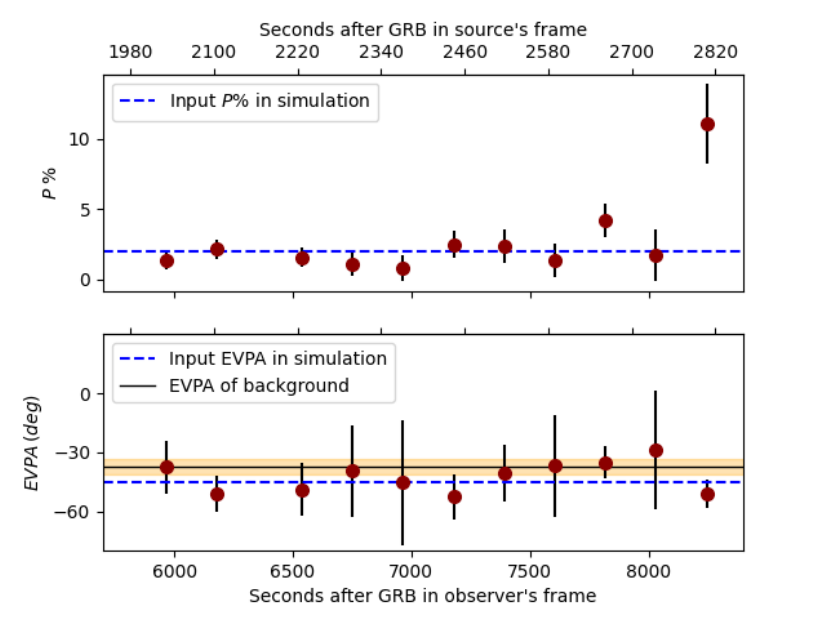}\\
    \end{tabular}
    \caption{Simulation measurements for a source with input polarization $P=2\%$. Each panel shows the case of a different input EVPA: $0,45,90,135^\circ$. Top of each panel: Measurements of polarization for each of the simulated frames with their 1$\,\sigma$ uncertainties. Bottom in each panel: EVPA measurements for each of the simulated frames with their 1$\,\sigma$ uncertainties, together with the measured EVPA and 1$\,\sigma$ uncertainty of the polarized sky for comparison. We note that the last exposure could not be measured in the two out of the four simulations (for input EVPAs of 0 and 45 degrees). EVPAs are shown interchangeably either in the (-90,90] or [0,180) range for clarity.}
    \label{fig:sims}
\end{figure*}

We present the time evolution of the degree of polarization ($P$) and the Electric Vector Position Angle (EVPA) of our observations in Fig.~\ref{fig:results}. Since polarization is a nonnegative quantity, measurements are biased toward higher values, especially those with low signal-to-noise ratio \citep[e.g.][]{Vaillancourt2006}. We present in Table~\ref{table:p_evpa} both the raw measurements and the debiased values for the polarization. Debiasing was performed according to \cite{Plaszczynski2014}. All the provided values in Table~\ref{table:p_evpa} and Fig.~\ref{fig:results} are corrected for instrumental polarization and ISP. Since the observations were made during morning twilight, we considered the possibility that the rise of the background affected our measurements (the morning sky is highly polarized). The stars in the RoboPol field cannot be confidently measured to compare how their polarization changes with time, allowing us to investigate this scenario. For this reason, we ran simulations as follows:
We made four fake images for each of our exposures by replacing the GRB source in our images with a mock source with polarization $P=2\%$, and a different value of EVPA for each of the four images: $0,45,90,135 ^\circ$. The FWHM and intensity of the source in each frame matched the FWHM and intensity of the corresponding real exposure. Then, for each of the fake frames, we conducted the analysis in the same way as for the real observations. We present the output of the simulations in Fig.~\ref{fig:sims}. It becomes obvious that the latter observations in all cases tend to be farther away from the real values than the former. Especially in the simulation with input $EVPA=0^\circ$ (top left of Fig.~\ref{fig:sims}), the measurements of EVPA tend to drift towards the background value. A similar drift seems to be apparent in the last measurements of the simulation with input $EVPA=45^\circ$ (top right of Fig.~\ref{fig:sims}). Finally, the latter polarization values of all simulations seem to be, on average, less accurate than the first ones. Based on the above, we conclude that the first five measurements are likely the only ones not affected by the polarized sky, since they are the only ones not affected by the background in all simulations. We present here all of our measurements, yet we highlight that the last six measurements of the time series are probably seriously affected by the high background.
Although RoboPol allows for background estimation for each of the four spots of the source to account for occasions when the background is polarized, in these particular images the source is fairly faint, the background high and the sky highly polarized ($\sim40\%$). Therefore, minor deviation of the background estimate from its true value is enough to give rise to the observed behavior. 
We note that for the first five measurements of the simulation, the scatter in the measured values in the simulations is similar to the one in the observations. Therefore, we postulate that the polarization and EVPA of the GRB are likely constant throughout these measurements. By combining them we get the debiased values $P=1.5\pm0.3$, $EVPA=8\pm6^\circ$.

Unlike the previous GRB observed with RoboPol at Skinakas \citep{King2014a}, where the polarization of GRB~131030A was found to be consistent with the interstellar polarization, we get a high-confidence (5 $\sigma$) detection of intrinsic GRB afterglow polarization. The degree of polarization measured in this work for GRB~210619B is rather typical for optical GRB afterglows in such scales \citep[e.g.][]{Covino2016}. According to the modeling of \cite{Oganesyan2021}, at the time of our observations the optical emission was dominated by the contribution of the forward shock. This level of linear polarization agrees with the long-thought idea that it arises primarily from synchrotron radiation. There are several different models that agree with this level of polarization. However, observations of temporal evolution of GRB polarization over a longer time period are more appropriate to put constraints and test the theory (see \cite{Gill2021} for a recent review).

\begin{table*}
\caption{Measurements of the degree of polarization and EVPA of the GRB. Time corresponds to the median time in the observer's frame of each 200 second exposure.}
\label{table:p_evpa}      
\centering          
\begin{tabular}{S[table-format=4.0] S[table-format=4.0]  S[table-format=-1.1,table-figures-uncertainty=2] S[table-format=-1.1,table-figures-uncertainty=2] S[table-format=-2.0,table-figures-uncertainty=2] S[table-format=-2.0,table-figures-uncertainty=2]}
\hline\hline
\multicolumn{2}{c}{Time since GRB (seconds)} & P  &  P$_{debiased}$ & {EVPA} & {EVPA}$_{debiased}$\\
       {Observer's frame} & {Source's frame} & \% & \% & $^\circ$ & $^\circ$ \\
\hline           \\         
\multicolumn{6}{c}{Likely not affected by the background}\\
\hline
    5967  & 2032  & 0.9\pm0.7 & 0.7\pm0.7 &37 \pm 21 &37 \pm 29 \\
    6177 & 2103  & 1.6\pm0.7 & 1.4\pm0.7 &-6 \pm 13 &-6 \pm 15 \\
    6537 & 2226  & 2.3\pm0.7 & 2.2\pm0.7 &22 \pm 9 &22 \pm 10 \\
    6751 & 2299  & 2.7\pm0.8 & 2.6\pm0.8 &2 \pm 8 &2 \pm 8 \\
    6964 & 2371  & 1.7\pm0.8 & 1.5\pm0.8 &14 \pm 14 &14 \pm 16 \\
    \hline
    \multicolumn{2}{c}{Average} & 1.6\pm0.3 & 1.5\pm0.3 &8 \pm 5 &8 \pm 6\\
    \hline  \\
    \multicolumn{6}{c}{Likely affected by the background}\\
    \hline
    7178 & 2444  & 4.1\pm.9 & 4.0\pm0.9 &-9 \pm 7 &-9 \pm 7 \\
    7391 & 2517  & 1.1\pm1.0 & 0.8\pm1.0 &5 \pm 27 &5 \pm 36 \\
    7604 & 2589  & 2.5\pm1.1 & 2.3\pm1.1 &-31 \pm 12 &-31 \pm 14 \\
    7818 & 2662  & 3.4\pm1.5 & 3.1\pm1.5 &-17 \pm 13 &-17 \pm 15 \\
    8031 & 2735  & 5.6\pm1.7 & 5.3\pm1.7 &2 \pm  8 &2 \pm 10 \\
    8245 & 2807  & 1.0\pm2.0 & 0.6\pm2.0 &-23 \pm 56 &-23 \pm 52 \\
\hline
\end{tabular}
\end{table*}

\begin{acknowledgements}
 We thank the anonymous referee and the editor Sergio Campana for providing useful comments that helped improved this manuscript. N.M, D.B., K.T., and K.K. acknowledge support from the European Research Council (ERC) under the European Union Horizon 2020 research and innovation program under grant agreement No 771282. K.T. acknowledges support from the Foundation of Research and Technology - Hellas Synergy Grants Program through the project POLAR, jointly implemented by the Institute of Astrophysics and the Institute of Computer Science. V.P. and S.R. were supported by the Hellenic Foundation for Research and Innovation (H.F.R.I.) under the "First Call for H.F.R.I. Research Projects to support Faculty members and Researchers and the procurement of high-cost research equipment grant" (Project 1552 CIRCE). V. P. acknowledges support from the Foundation of Research and Technology - Hellas Synergy Grants Program through project MagMASim, jointly implemented by the Institute of Astrophysics and the Institute of Applied and Computational Mathematics. D. R. A.-D. and J. A. acknowledge support by the Stavros Niarchos Foundation (SNF) and the Hellenic Foundation for Research and Innovation (H.F.R.I.) under the 2nd Call of “Science and Society” Action Always strive for excellence– “Theodoros Papazoglou” (Project Number: 01431).

\end{acknowledgements}

\bibliographystyle{aa}
\bibliography{bibliography}
\end{document}